\documentclass[preprint,twocolumn,5p]{elsarticle}
\pdfoutput=1
\usepackage{natbib}
\usepackage{blindtext, graphicx}
\usepackage{doi}
\usepackage[table]{xcolor}
\usepackage{paralist}
\usepackage{graphicx}
\usepackage{multirow}
\usepackage{color}
\usepackage[framed]{ntheorem}
\usepackage{framed}
\usepackage{tikz}
\usepackage{rotating}
\usepackage{pgfplots}
\usepackage{colortbl}
\usepackage{amsmath}
\usepackage{slashbox}

\usepackage{balance}
\usepackage{booktabs}
\usepackage{longtable}
\usepackage{listings}
\usepackage[lofdepth,lotdepth]{subfig}
\usepackage[font=small]{caption}
\usepackage{amsfonts}

\newcommand{\bi}{\begin{compactitem}}
\newcommand{\ei}{\end{compactitem}}
\newcommand{\be}{\begin{compactenum}}
\newcommand{\ee}{\end{compactenum}}
\newcommand{\overbar}[1]{\mkern 1.5mu\overline{\mkern-1.5mu#1\mkern-1.5mu}\mkern 1.5mu}

\expandafter\def\csname ver@subfig.sty\endcsname{} 
\usetikzlibrary{shadows}
\theoremclass{Lesson}
\theoremstyle{break}

\pgfplotsset{compat=1.14}
\begin{document}
%
\begin{frontmatter}
\title{Better Predictors for Issue Lifetime}  
\author{Mitch Rees-Jones\corref{cor1}}
\ead{mwreesjo@ncsu.edu}
\author{Matthew Martin\corref{cor1}}
\ead{mjmartin@colby.edu}
\author{Tim Menzies\corref{cor1}}
\ead{tim.menzies@gmail.com}
\address{Department of Computer Science, North Carolina State University, Raleigh, NC, USA}

\begin{abstract}

Predicting issue lifetime can help software developers, managers, and stakeholders 
effectively prioritize work, allocate development resources, and better understand 
project timelines. Progress had been made on this prediction problem, 
but prior work has reported low precision and high false alarms. The latest 
results also use complex models such as random forests that detract from their 
readability.

We solve both issues by using small, readable decision trees (under 20 lines long) and correlation feature selection to 
predict issue lifetime, achieving high precision and low false alarms (medians of 71\% and 13\% respectively). We also 
address the problem of high class imbalance within issue datasets - when local 
data fails to train a good model, we show that cross-project data can be used in 
place of the local data. In fact, cross-project data works so well that we argue 
it should be the default approach for learning predictors for issue lifetime.
\end{abstract}

\end{frontmatter}



\section{Introduction}


Predicting issue close   time
has multiple benefits for the developers, managers, and stakeholders. It helps  developers  prioritize work; helps managers  allocate resources and improve consistency of release cycles; and  helps  stakeholders understand changes in project timeliines and budgets. It is also useful to  predict issue close time  when an issue is created; e.g.
to  send a notification if it is predicted that the current issue is  an easy fix.

This paper  extends and improves  issue lifetime prediction methods
recently proposed 
by Kikas,  Dumas,  and  Pfahl (hereafter, KDP) at MSR'16~\cite{KDP}.
From KDPs recent paper, we will borrow two terms: 
an \emph{issue} is a bug report, and the time required to mark it   closed is the \emph{issue close time}. 

Our approach has similarities to that of KDP; however, we are not a exact replication of their study.
For example, 
KDP use Random Forests to build their models. Random
Forests are hard to read and reason from.
In our approach, we use Hall's CFS feature selector (described below) plus single decision tree learning 
yielded easily comprehensible trees (under 20 lines).
We find that this approach results in better predictive models.
KDP report precisions under 25\%, and false
alarm rates of over 60\%.
On the other hand, using CFS on single decision trees results in  
 precisions over 66\% 
and false alarms under 20\%. 

We also find that  
if local data fails to build a good model, then
cross-project data filtered by CFS can be used to build effective predictors. 
In fact,
cross-project data works so well, that we argue it should be the default
approach for learning predictors for issue lifetime.
 This success of cross-project learning for
 lifetime prediction was quite unexpected. The 
 cross-project results of this paper were achieved without the
data adjustments recommended in the defect prediction transfer
learning literature (see relevancy filtering~\cite{turhan2009relative,kocaguneli2015transfer} or   dimensionality
transform~\cite{nam2015heterogeneous}). 
That is, while transfer learning for some software analytics
tasks can be complex, there exist other tasks (such as predicting issue
lifetimes) where it is a relatively simple task.

 Overall, the  contributions of this paper are:
\bi
\item
A new ``high water mark'' in predicting issue close time;
\item
A new ``high water mark'' in cross-project prediction.
\item
A method for repairing poor local performance using cross-project learning.
\item
Evidence for that some software analytics tasks
allow for the very simple transfer of data between projects.
\item
A reproduction package containing all our scripts and data available online on GitHub\footnote{github.com/reesjones/issueCloseTime} and Zenodo\footnote{doi.org/10.5281/zenodo.197111}, which other researchers can use to repeat, improve, or even refute the claims of this paper.
\ei
This paper is organized as follows. 
The \emph{Background} section summarizes the current state of issue lifetime prediction. Next, in the \emph{Methods} section, we describe our learners and  data
(we  use issues, and code contributors come from GitHub and JIRA projects, from which we extracted ten issue datasets with a minimum, median, and maximum
number of  issues of 1,434, 5,266, 12,919 per dataset, respectively.
Our \emph{Results} section presents experimental results that defend three claims:
\bi
\item {\em Claim \#1}: Our predictors had low false alarms and higher precisions
than KDP.
\item {\em Claim \#2}: Cross-project learning works remarkably effectively
in this domain.
\item {\em Claim \#3}: Our predictors are easily comprehensible.
\ei


\section{Related Work}
  
 
Panjer~\cite{PanjerEclipse} explored  predicting the time to fix a bug using data  known at the beginning of a bug's lifetime. He used logistic regression models to achieve 34.9\% accuracy in classifying bugs as closing in 1.4 days, 3.4 days, 7.5 days, 19.5 days, 52.5 days, 156 days, and greater than 156 days.

Giger, Pinzger, and Gall \cite{GigerDecisionTrees}  used decision tree learning to 
make prediction models for issue close time for Eclipse, Gnome, and Mozilla bugs. 
They divided the time classes into thresholds of 1 day, 3 days, 1 week, 2 weeks, and
1 month, using static features for initial predictions that achieved a mean precision 
and recall of 63\% and 65\%. They also extended their models to include non-code 
features, which resulted in a 5-10\% improvement in model performance. Their 
predictions using non-code features were validated with 10-fold cross validation, 
meaning their train/test splits in the cross validation could have used an issue's 
data at a future point in time to predict its close time in the past (i.e. the leakage problem). Both of our local and cross-project methods avoid this potential conflation since all 
features used in our prediction are static, not temporal.

\begin{figure*}
\begin{center}
\includegraphics[width=7in]{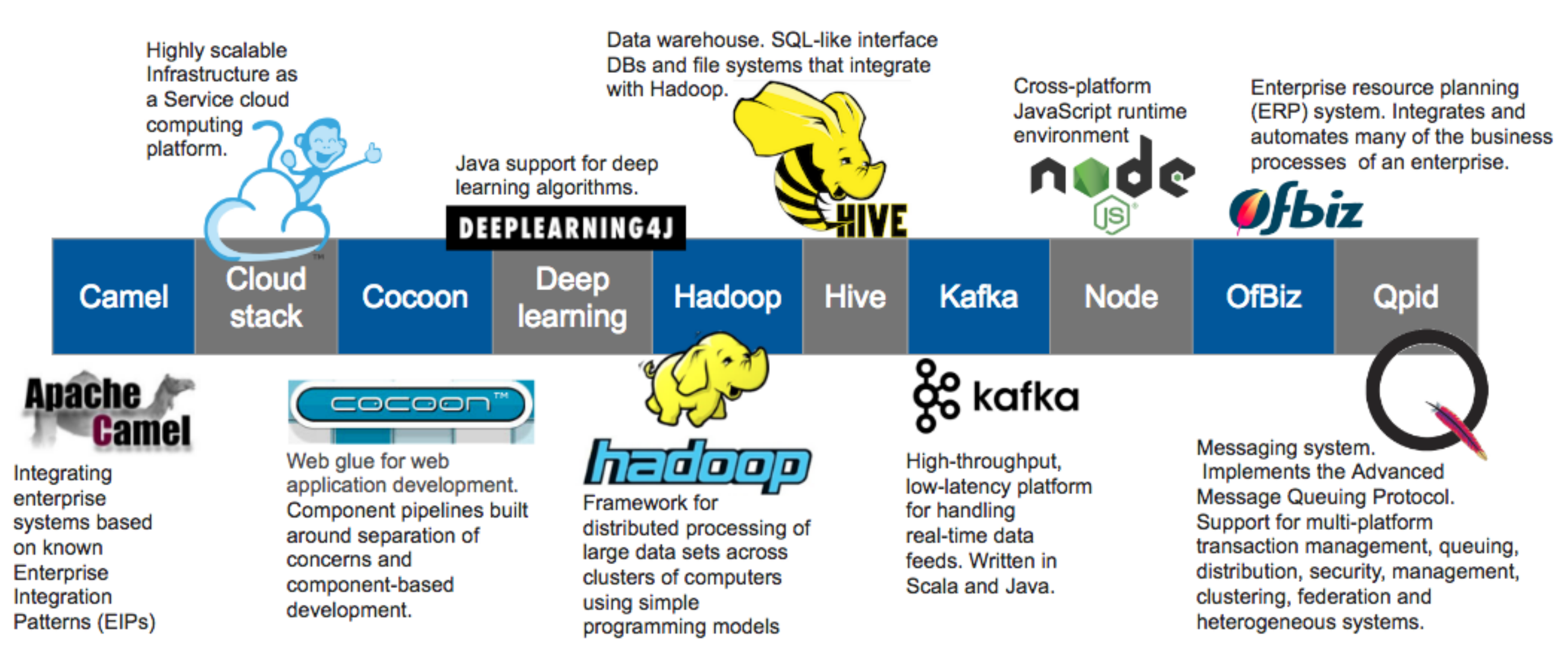}

\end{center}
\caption{The open-source projects that were used in this study. We extracted one issue lifetime dataset for each project listed above.}\label{fig:alldata}
\end{figure*}

Bhattacharya and Neamtiu \cite{BN} studied how existing models ``fail to validate on large projects widely used in bug studies''. In a comprehensive study, they find that the predictive power (measured by the coefficient of determination $R^2$) of existing models is 30-49\%. That study  found that there is no correlation between bug-fix likelihood, bug-opener reputation, and time required to close the bug for the three datasets used in their study.  Our results agree and disagree with this study. Like Bhattacharya and Neamtiu, we find that no single feature is always
most predictive of issue close time. That said, we do find that in different projects that different features
are highly predictive for issue close time. 

Guo, Zimmerman, Nagappan, and Murphy \cite{GuoWindows} evaluated the most predictive factors that affect bug fix time for Windows Vista and Windows 7 software bugs. Unlike Bhattacharya and Neamtiu's work \cite{BN}, they found that bug-opener reputation affected fix time; an issue with a high-reputation creator was more likely to get fixed. Bugs are also more likely to get fixed if the bug fixer is in the same team as or in geographical proximity of the bug creator. Guo et. al. conclude that the factors most important in bug fix time are social factors such as history of submitting high-quality bug reports and trust between teams interacting over bug reports. This conclusion of process metrics over product metrics is endorsed by \cite{ProcessMetrics}. 
Our results could be viewed as a partial replication of that study.
Like Guo et al. we find that non-code features are most predictive of issue close time (and this
is a {\em partial replication} since our findings come from different projects
to those used by Guo et al.). Also, our results extend those of Guo et al.
since we also check how well our preferred non-code features work in across 10 projects.

Marks, Zou, and Hassan \cite{MarksOSS} found that time and location of filing the bug report were the most important factors in predicting Mozilla issues, but severity was most important for Eclipse issues. Priority was not found to be an important factor for either Mozilla or Eclipse. Their models produced lower performance metrics (65\% misclassification rate) than subsequent work.

Zhang, Gong, and Versteeg \cite{ZhangKNN} reviewed the above work and concluded
that standard methods were insufficient to handle predicting issue lifetimes.
Hence, they developed an interesting, but intricate, new approach:
\bi
\item
A Markov-based method for predicting the number of bugs that will be fixed in the future;
\item
A method for estimating the total amount of time required to fix them;
\item
A kNN-based classification model for predicting a bug's fix time (slow or fast fix).
\ei
One reason
to prefer our approach  is that our methods can be implemented
with a very simple extension to current data mining tool kits. 
Zhang et al.'s approach  is  interesting, but we find that a very small change to standard
methods  (i.e. prepending learning with CFS) leads to 
comparable results. 
 Also, from our experience, the mathematical formalisms used in Zhang, et al. are difficult for some users to understand and thus gain deeper insight from. Therefore, we would prefer to use human-readable methods; e.g. data miners that learn single trees.

At MSR'16, Kikas, Dumas, and Pfahl (the KDP team)~\cite{KDP} built 
time-dependent models for issue close time prediction using Random Forests with a combination of static code
features,  and non-code features to predict issue close time with high performance. We regard this paper as the prior
state-of-the-art in predicting issue lifetime. Our results, reported below,
achieve similar performance figures to those of KDP, but we do so:
\bi
\item {\em  Using cross-project data.} This is an important extension to KDP since, as will
show, sometimes projects have very bad local data about project issues. In those
cases, our cross-project results would be preferred to the local-learning methods of KDP.
\item
{\em Using fewer features.}  KDP argue that certain
features are overall most important for predicting issue lifetime.
Our results suggest that their conclusion on ``most important feature''
need some adjustment since the ``most important'' features different from
project to project.
\item
{\em Using simpler learners.}
KDP used Random Forests which can be too large for humans to read. Our approach, that combines feature selection
with learning  a single decision tree, leads to very small, easily-readable models.
\ei
While we adopt much of the feature engineering of KDP, we do make some simplifications to their technique.
Rather than build 28 \emph{different} models for specific time periods in the issue's lifetime,
we only build five. We do this since, in our experience, the simpler the modeling, the more the 
commercial acceptance.


\section{Methods}

\subsection{Data}

Figure~\ref{fig:alldata} shows the ten projects used in this study, and Table~\ref{tab:projectStats} shows the number of issues and commits in each project when we collected the data.
These projects were selected by our industrial partners since
they use, or extend, software from these projects. The raw data dumps came in the form of commit data, issues, and code contributors from GitHub and JIRA projects, and we extracted ten issue datasets from them with a minimum of 1,434 issues, maximum of 12,191, and median of 5,267 issues per dataset. We also created a ``combined" dataset, which aggregated all instances from the ten datasets into one, for a total of eleven datasets used in the study.

In raw form, our data consisted of  sets of JSON files for each  repository, each file containing one type of data regarding the software repository (issues, commits, code contributors, changes to specific files), with one JSON file joining all of the data associated with a commit together: the issue associated with a commit, commit time, the magnitude of the commit, and other information. In order to extract data specific to issue close time, we did some preprocessing and feature extraction on the raw datasets pertaining to issue close time.

\begin{table}[]
{\small \centering
\begin{tabular}{@{}ccc@{}}
\toprule
Project                           & Number of Issues & Number of Commits \\ \midrule
\multicolumn{1}{c|}{camel}        & 5056             & 1335              \\
\multicolumn{1}{c|}{cloudstack}   & 1551             & 2257              \\
\multicolumn{1}{c|}{cocoon}       & 2045             & 1135              \\
\multicolumn{1}{c|}{deeplearning} & 1434             & 1113              \\
\multicolumn{1}{c|}{hadoop}       & 12191            & 1588              \\
\multicolumn{1}{c|}{hive}         & 5648             & 859               \\
\multicolumn{1}{c|}{kafka}        & 1732             & 400               \\
\multicolumn{1}{c|}{node}         & 6207             & 786               \\
\multicolumn{1}{c|}{ofbiz}        & 6177             & 1530              \\
\multicolumn{1}{c|}{qpid}         & 5475             & 1422              \\ \bottomrule
\end{tabular}}
\caption{The total number of issues and commits per project.}
\label{tab:projectStats}
\end{table}

\begin{table}[!t]\small
  \begin{tabular}{p{3cm}|p{5cm}}
  
    \textbf{ Feature name}               & \textbf{ Feature Description} \\ \hline
    \emph{issueCleanedBodyLen}          & The number of words in the issue title and description. For JIRA issues, this is the number of words in the issue description and summary \\ 
    \emph{nCommitsByCreator}            & Number of commits made by the creator of the issue in the 3 months before the issue was created \\  
    \emph{nCommitsInProject}            & Number of commits made in the project in the 3 months before the issue was created \\  
    \emph{nIssuesByCreator}             & Number of issues opened by the issue creator in the 3 months before the issue was opened \\  
    \emph{nIssuesByCreatorClosed}       & Number of issues opened by the issue creator that were closed in the 3 months before the issue was opened \\ 
    \emph{nIssuesCreatedInProject}      & Number of issues opened in the project in the 3 months before the issue was opened \\ 
    \emph{nIssuesCreatedIn-\newline ProjectClosed}& Number of issues in the project opened and closed in the 3 months before the issue was opened \\\hline
    \emph{timeOpen } & Close time of the issue (target class).
  \end{tabular}
  \caption{Features used in this study (prior to feature selection). A \emph{creator} is the person who opened the issue. A \emph{project} is a software repository that has associated issue and commit data.}
  \label{tab:features}
\end{table}

Our feature engineering is based on KDP's study, but we diverge from their approach in a few ways. We chose to make our model simpler and smaller with the hopes of   making   models easier to read and understand. We also do not use any of KDPs temporal features so that we can make predictions at issue creation time. For these reasons, of the 21 features used by KDP, we first eliminated all features which:
\bi
\item
Were not available in our raw data,
\item
Could not calculated. For example, KDP do not specify how to calculate their \emph{textScore} feature,
\item
Needed data generated after issue creation.
\ei
This brought our feature list down to just seven features, as detailed in Table \ref{tab:features}.
Note that our target class was \emph{timeOpen}.


Finally, regarding our methods for handling data, the following two points are critical:
\be
\item
One issue in preparing this data is that a small number of issues were
 \emph{sticky}. A sticky issue is one which was not yet closed at the time of data collection. In the KDP paper, sticky issues were handled by approximating the close time to be a chosen set date in the future. KDP's method for doing that is innovative,
 but somewhat subjective. Hence, we simply omitted sticky issues
 from our data sets. 
 \item
 In this kind of study, it would be a methodological error to build models by using {\em future} data points to predict for {\em past} events (i.e. the leakage problem in machine learning). In this study, we emphasize that because {\em none} of our features are time-dependent (all features are known at issue creation time), there is no risk of leakage.
 \ee

\subsection{Selecting Target Classes}
In the issue lifetime prediction literature, there are
several ways to handle the  learning target: 
\bi
\item
Bhattacharya and Neamtiu use multivariate regression to predict an exact fix time~\cite{BN}. Note that such regression models output one continuous value.
\item
Both Panjer and Marks et al. build one classifier that predicts for N classes~\cite{PanjerEclipse,MarksOSS}.

\item
KDP, Zhang et. al., and Giger et. al. build one binary classifier for N goals~\cite{KDP,ZhangKNN,GigerDecisionTrees}.
This ``N-binary'' approach 
is standard practice   when using 
support vector machines~\cite{yu2016read}.
\ei
For two reasons, we adopt this ``N-binary'' approach.
Firstly, as KDP is the current state-of-the-art, we adopt their approach.
Accordingly, instead
of asking what will be the issue close time, we
discretized the issue report time into five lifetimes:
{\em day, week, 2weeks, month,} and {\em 3months};
then built five predictors for the following two-class problems:
\bi
\item Predictor1: closes in 1 day vs more than 1 day;
\item Predictor2: closes in 1 week vs more than one week;
\item Predictor3: closes in 2 weeks vs more than 2 weeks;
\item Predictor4: closes in 1 month vs more than 1 month;
\item Predictor5: closes in 3 months vs more than 3 months.
\ei
Our second reason for using ``N-binary'' classifier is that
it fit the needs of our industrial partners. 
We  work  with  a  group  of  developers in Raleigh, NC that attend a monthly ``open issues" report with their management. The sociology of that meeting is  that  the fewer the  open  issues,  the less management  will interfere  with  particular  projects.  In  this  context,  developers are  motivated  to  clear  out  as  many  issues  as  possible  before that meeting. Therefore, the question these developers want to know is what issues might be closed well before that monthly meeting (i.e. can this issue be closed in one week or two?).

Generalizing from the experience with our industrial partners, 
 we  say  that ``N-binary''  learning  is  appropriate  when  the  local  user  population  has  ``activity  thresholds'' where new activity is required if some measure reaches some threshold point. In such a context, the more general and harder question of ``is it
 class X or Y or Z'' can be replaced by the simpler and more specific question of ``is it a result that crosses our thresholds?''

\begin{figure}[!t]
  \centering
  \begin{tikzpicture}
  	\begin{axis}[ybar stacked,
  	    xtick=data,
  	    symbolic x coords={1,7,14,30,90,180,365,1000},
  	   legend style={at={(0.5,-0.20)},anchor=north,legend columns=4},
  	   xlabel={Time class (before $n$ days)},
  	   ylabel={Relative size of class},
  	]
  	\addplot coordinates 
          {(1,0.3405) (7,0.2132) (14,0.0722) (30,0.0815) (90,0.1463) (180,0.3205) (365,1.0000) (1000,0.2917)};
      \addplot coordinates 
          {(1,1.0000) (7,0.6956) (14,0.1537) (30,0.1629) (90,0.2024) (180,0.0989) (365,0.0365) (1000,0.0091)};
      \addplot coordinates 
          {(1,0.0812) (7,0.0591) (14,0.0208) (30,0.0292) (90,0.0552) (180,0.0338) (365,0.0481) (1000,1.0000)};
      \addplot coordinates 
          {(1,1.0000) (7,0.2301) (14,0.0817) (30,0.0774) (90,0.0742) (180,0.0419) (365,0.0344) (1000,0.0011)};
      \addplot coordinates 
          {(1,0.0073) (7,0.0116) (14,0.0194) (30,0.0718) (90,0.3161) (180,0.3964) (365,0.3870) (1000,1.0000)};
      \addplot coordinates 
          {(1,0.0118) (7,0.0144) (14,0.0380) (30,0.1166) (90,0.6887) (180,0.8873) (365,0.9436) (1000,1.0000)};
      \addplot coordinates 
          {(1,1.0000) (7,0.6798) (14,0.2266) (30,0.2069) (90,0.2221) (180,0.0906) (365,0.0710) (1000,0.1178)};
      \addplot coordinates 
          {(1,1.0000) (7,0.7410) (14,0.2157) (30,0.1869) (90,0.2281) (180,0.1037) (365,0.0729) (1000,0.0066)};
      \addplot coordinates 
          {(1,1.0000) (7,0.7716) (14,0.3083) (30,0.3149) (90,0.3809) (180,0.3096) (365,0.2647) (1000,0.7267)};
      \addplot coordinates 
          {(1,0.0703) (7,0.0654) (14,0.0292) (30,0.0620) (90,0.1946) (180,0.2983) (365,0.1855) (1000,1.0000)};
    \legend{camel,cloudstack,cocoon,deeplearning,hadoop,hive,kafka,node,ofbiz,qpid}
  	\end{axis}
  \end{tikzpicture}
  \caption{Class distribution for the \emph{timeOpen} feature for each of the ten issue lifetime datasets (plus one dataset combining all 10) used in the study. Many issues were closed between 0 and 1 day, and between 365 and 1000 days.}
  \label{fig:classFrequencies}
\end{figure}
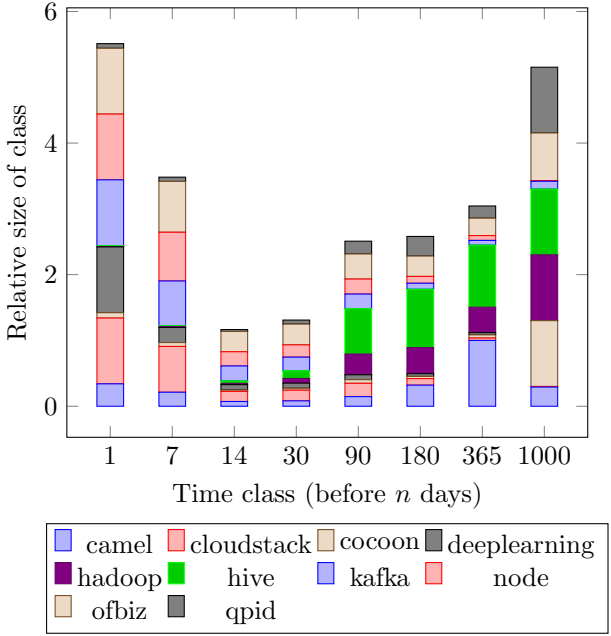

\subsection{Class Re-Balancing}
The \emph{timeOpen} class distribution for each dataset is shown in Figure \ref{fig:classFrequencies}. Note that many issues were closed within a day or before 7 days, as well as between 365 and 1000 days, while very few issues were closed between 14 and 90 days. 

 For some of our datasets on certain time thresholds, the minority to majority class ratio was over 300:1, which created difficulty for the learners to train themselves effectively. To handle this problem of highly imbalanced class
distributions, we tried applying SMOTE (Synthetic Minority Over-Sampling~\cite{SMOTE})
which an is oversampling technique for equalizing class distributions. 
The results of SMOTEing were inconclusive since no statistically significant difference was detected between our SMOTEd and non-SMOTEd
results. Accordingly, this report makes no further mention of SMOTE.

\subsection{Feature Subset Selection}

It can be surprisingly useful to ignore some features in a data set.
Hall et al.~\cite{hall1999correlation} note that the effect of
 such {\em feature selection} can be quite dramatic, particularly
for decision tree learners.
If a data set has, say, four features
that are strongly associated with the target class,
a binary decision tree learner would quickly use those features
to divide the data $2^4=16$ ways.
This means that subsequent learning, at lower levels of
tree, can only reason about $1/16$-th of the data.
On the other hand, if those four features are strongly
associated with each other, feature selection would
remove all but one of them prior to learning.
The decision tree learner would then use that single 
feature
to divide the data twice, after which point subsequent
learning at lower levels of tree could reason better
(since it can access more data).

KDP explored some limited feature selection from their data.
Their method was an ``arity=1'' technique that commented on the
value of each feature, when explored in isolation. Current work
in machine learning on feature selection uses ``arity$>$1'' methods
that report the value of {\em sets} of features.
Accordingly,
this section reports the effect of such ``arity$>$1'' feature
selectors on issue close-time data.

Specifically, Hall's
CFS feature selector~\cite{hall1999correlation} was used
to determine which features were most important.
Unlike some other feature selectors (e.g. Relief, InfoGain), 
CFS evalautes and hence ranks feature subsets rather than 
individual features. 
and hence CFS is based on the heuristic
that ``good feature subsets contain features highly correlated with the classification, yet uncorrelated to each other''. Using this heuristic,
CFS performs a best-first search (with a horizon of five\footnote{(1) The initial ``frontier''
is  all sets containing
one different feature. (2) The frontier of size $n$ (initially $n=1$) is sorted according to 
{\em merit} and the best item is grown to all sets
of size {\em n+1} containing the best item from the last
frontier. Go to step (3). Halt when last five frontiers
have not seen an improvement in $merit$. On halt, return
the best subset seen so far.}) to discover interesting sets of features

\noindent 
Hall et al. scores each feature subsets as follows:
\[
\mathit{merit}_s = \frac{k\overbar{r_{\mathit{cf}}}}{\sqrt{k+k(k-1)\overbar{r_{\mathit{ff}}}}}
\]
where:
\bi
\item
$\mathit{merit}_s$ is the value of some subset $s$ of the
features containing $k$ features; 
\item
$r_{\mathit{cf}}$ is a score describing the connection of that feature
set to the class;
\item
and $r_{\mathit{ff}}$ is the mean score of the feature to feature
connection between the items in $s$.
\ei
Note that for this to be maximal, $r_{\mathit{cf}}$  must be large
and $r_{\mathit{ff}}$ must be small. That is, features have to connect
more to the class than each other. 

The above equation is actually Pearson's correlation  where
all variables have been standardized.  To be applied
for discrete class learning (as done by KDP and this paper),
Hall et al. employ the Fayyad Irani discretizer~\cite{FayIra93Multi} then apply the following
entropy-based measure to infer $r$ (the degree of associations
between  discrete sets $X$ and $Y$):

\[
r_{\mathit{xy}}=2\times \left[ \frac{H(x) + H(y) - H(x,y)}{H(y)+H(x)} \right]
\]
where $H$ is the standard information gain measure used in 
decision tree learning.

\subsection{Decision Tree Learning}
 Decision trees are standard classification models that use the concept of entropy in information theory to partition data into classes in a way that either minimizes entropy or maximizes homogeneity in each partition. Decision tree learners attempt to predict the value of the target variable (in this case, \emph{timeOpen}) by recursively partitioning the data on features that most decrease the information entropy of each partition, until a stopping criterion is reached (such as number of instances in a partition being less than a chosen threshold).

Decision tree learners by nature are prone to overfitting the training set, usually when the stopping criterion is not aggressive enough, since a decision tree can be built to perfectly classify the training set. Overfitting can be avoided with a number of approaches, by setting a minimum number of instances per partition, which stops partitioning when the partition's size is less than the threshold size, or by pre- or post-pruning the tree, which is the process of replacing subtrees with single leaf nodes when doing so doesn't significantly increase the error rate.

We used the C4.5 decision tree learner, using an aggressive pruning parameter $M$ to stop partitioning the tree when the partition size was less than $M$, where we defined $M$ as:

\begin{equation}\label{eq:m25}
M = \frac{\text{number of instances in the dataset}}{25}
\end{equation}

The ``magic number'' of 25 was set after a pre-study that tried 
values of 100, 50, 25, 12, 6, and 3, and reported no significant
differences in performances between 25 and 3. 
Note that this definition of $M$ means that no sub-tree of our data
will ever be built from small samples of our data.

\subsection{Model Generation and Evaluation Loop}

We used the open-source data mining tool WEKA \cite{Weka}, developed at the Machine Learning Group at the University of Waikato, for all of our data mining operations, including dataset filtering, model generation, and model evaluation. We generated our models as follows:
\bi
\item
First, we first split each dataset into five datasets, one for each of our chosen prediction thresholds (1, 7, 14, 30, and 90 days).
\item
Next we applied correlation-based feature subset selection (CFS) on each data set to find what features are relevant to
each data set.
\item
Finally, we  built C4.5 decision trees on the resulting filtered and oversampled datasets. Note that, within WEKA, C4.5 is called J48 (short for ``Java port of C4.5 release 8'').
\ei
The resulting models were evaluated  in one of two ways:
\bi
\item Local learning with cross-val; i.e. performing stratified 10-fold cross validation where, 10 times, we test on 10\% of the data while training on the other
90\%.
\item
Cross-project learning with round-robin; i.e. start with $N=10$ projects,
for each project $p \in N$, train on $N-p$ then test on project $p$. 
\ei
Note that,     the round-robin studies were repeated
for each target time period. For example, when
trying to predict for ``within 1 week'' vs ``more than one week''
in project $p$, we only collected ``within 1 week'' vs ``more than one week'' data in the other $N-p$ data sets.

As shown below, in some cases the local learning generated
poor results. For all those cases, much better
results were seen using round-robin. Hence, for predicting issue
lifetimes,
we recommend the use of the round-robin approach.

\subsection{Performance Measures}

Precision, recall, and false alarms are three  performance measures for binary classification problems, where a data point is classified as ``selected'' or ``not selected'' by the model:

$$prec = \frac{TP}{TP + FP},\ recall = \frac{TP}{TP + FN},\ pf=\frac{FP}{TN+FP}$$

Here,  $\{TP,\ FP,\ TN,\ FN\}$ are the count of true positives, false positives, true negatives, false negatives, respectively.


\section{Results}

This section offers results that support the claims made
in the introduction:
\bi
\item {\em Claim \#1}: Our predictors had lower false alarms and higher precisions
than KDP.
\item {\em Claim \#2}: Cross-project learning works remarkably effectively
in this domain.
\item {\em Claim \#3}: Our predictors are easy to read \& comprehend.
\ei

\subsection{Claim \#1: Better Precisions and False Alarms}

Table~\ref{tab:comparison} shows   mean results averaged
over all data sets for this paper and KDP.  As can be seen:
\bi
\item KDP has slightly better recall results (pd);
\item Our methods have much higher precision (median 71\%);
\item Our methods have much lower false alarms (median 13\%);
\ei
In our engineering judgement, our false alarm and precision
results more than compensate with the slightly lower recalls.

\begin{table}[!t]
{\small
\centering
\begin{tabular}{c|ccc|ccc|}
               & \multicolumn{3}{c|}{Round Robin}  &\multicolumn{3}{c|}{KDP} \\
Time threshold & prec       & recall       & pf       & prec    & recall     & pf    \\ \hline
1 day          & .58        & .19      & .02      & N/A     & N/A    & N/A   \\ \cline{1-1}
7 days         & .66        & .68      & .12      & .26     & .82    & .63   \\ \cline{1-1}
14 days        & .71        & .73      & .13      & .13     & .80    & .64   \\ \cline{1-1}
30 days        & .77        & .71      & .12      & .16     & .80    & .65   \\ \cline{1-1}
90 days        & .78        & .73      & .17      & .25     & .79    & .64   \\ \hline
\end{tabular}}
\caption{Comparison between the performances of our Round Robin approach and KDP's results. We achieve better precisions and false positive rates, but slightly lower recall. Note that the N/As for KDP's 1-day time threshold exist because their time-dependent features do not allow them to predict the 1-day threshold from 0 days.}
\label{tab:comparison}
\end{table}
\definecolor{carmine}{rgb}{0.59, 0.0, 0.09}
\begin{table}[ht]
\renewcommand{\baselinestretch}{.95} 
\scriptsize
\centering
\begin{tabular}{cc|ccc|ccc}
\multicolumn{2}{l|}{}            
 & \multicolumn{3}{c|}{Crossval} & \multicolumn{3}{c}{Round Robin} \\
 & & \multicolumn{3}{c|}{  (local learning)} & \multicolumn{3}{c}{ (cross-project)} \\
Dataset                       & \begin{tabular}[c]{@{}c@{}}Time\\ Class\end{tabular} & {\em prec}    & {\em recall}    & {\em pf}    & {\em prec}     & {\em recall}     & {\em pf}    \\ \hline
\multirow{5}{*}{camel}        & 1                                                    & 65        & 40      & 4      & 56         & \cellcolor{carmine!25}{23}       & 3       \\ \cline{2-2}
                              & 7                                                    & 74        & 70      & 7      & 66         & 70       & 12      \\ \cline{2-2}
                              & 14                                                   & 73        & 78      & 10      & 74         & 70       & 11      \\ \cline{2-2}
                              & 30                                                   & 82        & 80      & 7      & 77         & 74       & 12      \\ \cline{2-2}
                              & 90                                                   & 89        & 70      & 5      & 77         & 81       & 21      \\ \hline
\multirow{5}{*}{cloudstack}   & 1                                                    & 66        & 60      & 24      & 57         & \cellcolor{carmine!25}{18}       & 2       \\ \cline{2-2}
                              & 7                                                    & 76        & 93      & \cellcolor{carmine!25}{77}      & 65         & 66       & 11      \\ \cline{2-2}
                              & 14                                                   & 81        & 96      & \cellcolor{carmine!25}{83}      & 70         & 71       & 11      \\ \cline{2-2}
                              & 30                                                   & 85        & 100     & \cellcolor{carmine!25}{100}       & 78         & 65       & 9       \\ \cline{2-2}
                              & 90                                                   & 94        & 100     & \cellcolor{carmine!25}{100}       & 74         & 75       & 20      \\ \hline
\multirow{5}{*}{cocoon}       & 1                                                    & \cellcolor{carmine!25}{0}         & \cellcolor{carmine!25}{0}       & 0       & 60         & \cellcolor{carmine!25}{17}       & 2       \\ \cline{2-2}
                              & 7                                                    & \cellcolor{carmine!25}{0}         & \cellcolor{carmine!25}{0}       & 0       & 66         & 68       & 13      \\ \cline{2-2}
                              & 14                                                   & \cellcolor{carmine!25}{0}         & \cellcolor{carmine!25}{0}       & 0       & 71         & 75       & 13      \\ \cline{2-2}
                              & 30                                                   & 53        & 96      & 14      & 77         & 71       & 11      \\ \cline{2-2}
                              & 90                                                   & 67        & 95      & 11      & 82         & 64       & 12      \\ \hline
\multirow{5}{*}{deeplearning} & 1                                                    & 78        & 77      & \cellcolor{carmine!25}{41}      & 51         & \cellcolor{carmine!25}{17}       & 3       \\ \cline{2-2}
                              & 7                                                    & 80        & 100     & \cellcolor{carmine!25}{100}       & 65         & 66       & 11      \\ \cline{2-2}
                              & 14                                                   & 86        & 100     & \cellcolor{carmine!25}{100}       & 70         & 73       & 12      \\ \cline{2-2}
                              & 30                                                   & 91        & 100     & \cellcolor{carmine!25}{100}       & 77         & 67       & 10      \\ \cline{2-2}
                              & 90                                                   & 96        & 100     & \cellcolor{carmine!25}{100}       & 76         & 77       & 19      \\ \hline
\multirow{5}{*}{hadoop}       & 1                                                    & \cellcolor{carmine!25}{0}         & \cellcolor{carmine!25}{0}       & 0       & 57         & \cellcolor{carmine!25}{22}       & 4       \\ \cline{2-2}
                              & 7                                                    & \cellcolor{carmine!25}{0}         & \cellcolor{carmine!25}{0}       & 0       & 66         & 70       & 18      \\ \cline{2-2}
                              & 14                                                   & \cellcolor{carmine!25}{0}         & \cellcolor{carmine!25}{0}       & 0       & 70         & 81       & 22      \\ \cline{2-2}
                              & 30                                                   & \cellcolor{carmine!25}{0}         & \cellcolor{carmine!25}{0}       & 0       & 76         & 80       & 20      \\ \cline{2-2}
                              & 90                                                   & \cellcolor{carmine!25}{32}        & \cellcolor{carmine!25}{2}      & 0       & 80         & 83       & 24      \\ \hline
\multirow{5}{*}{hive}         & 1                                                    & \cellcolor{carmine!25}{0}         & \cellcolor{carmine!25}{0}       & 0       & 61         & \cellcolor{carmine!25}{15}       & 2       \\ \cline{2-2}
                              & 7                                                    & \cellcolor{carmine!25}{0}         & \cellcolor{carmine!25}{0}       & 0       & 68         & 67       & 13      \\ \cline{2-2}
                              & 14                                                   & 52        & 35      & 1       & 73         & 71       & 13      \\ \cline{2-2}
                              & 30                                                   & \cellcolor{carmine!25}{0}         & \cellcolor{carmine!25}{0}       & 0       & 80         & 69       & 10      \\ \cline{2-2}
                              & 90                                                   & 62        & 53      & 9       & 81         & 76       & 16      \\ \hline
\multirow{5}{*}{kafka}        & 1                                                    & 63        & 48      & 17      & 58         & \cellcolor{carmine!25}{16}       & 2       \\ \cline{2-2}
                              & 7                                                    & 78        & 83      & \cellcolor{carmine!25}{43}      & 65         & 66       & 11      \\ \cline{2-2}
                              & 14                                                   & 81        & 90      & \cellcolor{carmine!25}{56}      & 69         & 73       & 12      \\ \cline{2-2}
                              & 30                                                   & 83        & 97      & \cellcolor{carmine!25}{76}      & 76         & 69       & 10      \\ \cline{2-2}
                              & 90                                                   & 91        & 98      & \cellcolor{carmine!25}{71}      & 81         & 62       & 11      \\ \hline
\multirow{5}{*}{node}         & 1                                                    & 56        & \cellcolor{carmine!25}31      & 16      & 60         & \cellcolor{carmine!25}{21}       & 2       \\ \cline{2-2}
                              & 7                                                    & 69        & 95      & \cellcolor{carmine!25}{89}      & 64         & 55       & 7       \\ \cline{2-2}
                              & 14                                                   & 76        & 100     & \cellcolor{carmine!25}{100}     & 68         & 55       & 7       \\ \cline{2-2}
                              & 30                                                   & 84        & 100     & \cellcolor{carmine!25}{100}     & 74         & 56       & 7       \\ \cline{2-2}
                              & 90                                                   & 93        & 100     & \cellcolor{carmine!25}{100}     & 69         & 69       & 18      \\ \hline
\multirow{5}{*}{ofbiz}        & 1                                                    & \cellcolor{carmine!25}{0}         & \cellcolor{carmine!25}{0}       & 0       & 63         & \cellcolor{carmine!25}{21}       & 2       \\ \cline{2-2}
                              & 7                                                    & 54        & 43      & 27      & 66         & 79       & 12      \\ \cline{2-2}
                              & 14                                                   & 56        & 70      & \cellcolor{carmine!25}{57}      & 73         & 78       & 10      \\ \cline{2-2}
                              & 30                                                   & 62        & 87      & \cellcolor{carmine!25}{77}      & 79         & 72       & 9       \\ \cline{2-2}
                              & 90                                                   & 67        & 100     & \cellcolor{carmine!25}{100}     & 83         & 64       & 9       \\ \hline
\multirow{5}{*}{qpid}         & 1                                                    & \cellcolor{carmine!25}0         & \cellcolor{carmine!25}0       & 0       & 60         & \cellcolor{carmine!25}16        & 2       \\ \cline{2-2}
                              & 7                                                    & \cellcolor{carmine!25}0          & \cellcolor{carmine!25}0        & 0       & 66         & 71       & 14      \\ \cline{2-2}
                              & 14                                                   & \cellcolor{carmine!25}0          & \cellcolor{carmine!25}0        & 0       & 70         & 78       & 16      \\ \cline{2-2}
                              & 30                                                   & \cellcolor{carmine!25}0          & \cellcolor{carmine!25}0        & 0       & 74         & 82       & 17      \\ \cline{2-2}
                              & 90                                                   & 53        & \cellcolor{carmine!25}19       & 5       & 79         & 75       & 18      \\  \cline{2-8}
\end{tabular}
\caption{Median predictive performance of each model created. Each row corresponds to the performance statistics of a dataset split by a certain time threshold. Cells marked with \textcolor{carmine}{{\bf red}}
indicate ``bad'' results; i.e. false alarms over 33\%
or precision or recall results under 33\%.}\label{tab:results}
\end{table}

\subsection{Claim \#2: Cross-Project Learning Works Well}
 
Table~\ref{tab:results} show median results for
precision, recall, and false alarm seen in our local learning
and round-robin experiments.  Cells marked with \textcolor{carmine}{{\bf red}} show ``bad'' results; i.e. very
low precisions or low recalls or high false alarms.
Three things to note about Table~\ref{tab:results} are:
\bi
\item
In many cases, the local results have many ``bad'' results. 
This result explains many of the results described in {\em Releated Work}; i.e. learning issue lifetimes is hard using just data rom
one project.
\item
With one exception, the cross-project results are not ``bad'';
i.e. cross-project learning performs very well for lifetime prediction.
\item
The one exception is predicting for issues that close in 1 day.
By its very nature, it is a challenging task since it relies
on a very small window for data collection. Hence, it is not
surprising that even our best round-robin learning scheme
has ``bad'' performance for this hard task.
\item
Interestingly, just because local learning has problems with
a data set does not mean that that cross-project learning
is also  challenged.  Table~\ref{tab:results} shows that
cross-project learning usually works very well.
\ei
The exception to this last point are the  cross-project recalls for {\em node}. These are quite low:  often 50\%  to 60\%. That aid, the the precisions for cross-project {\em node} are respectable
and the false alarms for cross-project {\em node} are far superior
to the local learning {\em node} results.

\subsection{Claim \#3: Our Models Are Easy to Explain} 
 
Compared to KDP, our results are easier to explain
to business users:
\bi
\item
Since we do use single-tree
decision tree learning rather than  Random
Forests (as done by KDP), it is not necessary to browse
across an ensemble of trees in a forest to understand our models.

\item
Since we use the early stopping rule of Equation~\ref{eq:m25},
our trees are  very small in size (median of 9 nodes in round-robin).
Figure \ref{fig:trees} shows two trees learned during the round-robin
when predicting for issues that close in 7 or 14 days for
{\em hadoop} or {\em camel}.  Each tree is six lines 
long. No other tree learned in this study was  more than 20 lines long.
\item
Since we use feature selection, our data miner has fewer
features from which they can learn trees. Accordingly,
our trees contain fewer concepts.
\ei
\begin{figure}[!t]
\renewcommand{\baselinestretch}{0.75}
  
{\scriptsize \begin{verbatim}
HADOOP-7:
       nissuescreatedinprojectclosed <= 33     :7  
       nissuescreatedinprojectclosed > 33
       |   nissuescreatedinprojectclosed <= 199 
       |   |   issuecleanedbodylen <= 27       :not7  
       |   |   issuecleanedbodylen > 27        :7  
       |   nissuescreatedinprojectclosed > 199 :not7 


CAMEL-14:
       nIssuesCreatedInProjectClosed <= 12 :14  
       nIssuesCreatedInProjectClosed > 12
       |   nCommitsInProject <= 596
       |   |   nCommitsInProject <= 524    :not14  
       |   |   nCommitsInProject > 524     :14  
       |   nCommitsInProject > 596         :not14  
\end{verbatim}}
\renewcommand{\baselinestretch}{1}
 \caption{Learned Decision Trees. These are nested if-then-else statements. For example, the last line of the the above is part of a branch saying the following kind of issue will not
 close in 14 days: {\em if nIssuesCreatedInProjectClosed is over 12 and if nCommitsInProject is over 596}.}\label{fig:trees}
 \end{figure}
 \begin{table}[!t]
\scriptsize
  \centering
  \begin{tabular}{r|c@{~}c@{~}c@{~}c@{~}c@{~}c@{~}c@{~}c@{~}c@{~}c@{~}c@{~}|c}
      \backslashbox{Feature name}{Dataset name} & \begin{turn}{90}node\end{turn} & \begin{turn}{90}hive\end{turn} & \begin{turn}{90}camel\end{turn} & \begin{turn}{90}kafka\end{turn} & \begin{turn}{90}ofbiz\end{turn} & \begin{turn}{90}combined\end{turn} & \begin{turn}{90}qpid\end{turn} & \begin{turn}{90}cloudstack\end{turn} & \begin{turn}{90}deeplearning\end{turn} & \begin{turn}{90}hadoop\end{turn} & \begin{turn}{90}cocoon\end{turn} & \begin{turn}{90}Total\end{turn} \\ \hline
      nCommitsInProject            & $\circ$ & $\circ$ & $\circ$ & $\circ$ & $\circ$ & ~        & ~        & $\circ$ & $\circ$ & $\circ$ & ~        & 8 \\
      nIssuesCreatedInProjectClosed& $\circ$ & $\circ$ & $\circ$ & $\circ$ & $\circ$ & $\circ$ & $\circ$ & ~        & ~        & ~        & $\circ$ & 8 \\
      nIssuesCreatedInProject      & $\circ$ & $\circ$ & $\circ$ & $\circ$ & ~        & $\circ$ & $\circ$ & $\circ$ & ~        & $\circ$ & ~        & 8 \\
      nIssuesByCreatorClosed       & $\circ$ & $\circ$ & ~        & $\circ$ & $\circ$ & ~        & $\circ$ & ~        & ~        & ~        & ~        & 5 \\
      nCommitsByCreator            & $\circ$ & $\circ$ & $\circ$ & ~        & ~        & $\circ$ & ~        & ~        & ~        & ~        & ~        & 4 \\
      issueCleanedBodyLen          & $\circ$ & ~        & ~        & ~        & $\circ$ & ~        & ~        & ~        & $\circ$ & ~        & ~        & 3 \\
      nIssuesByCreator             & $\circ$ & ~        & ~        & ~        & ~        & ~        & ~        & ~        & ~        & ~        & ~        & 1 \\ \hline
      Total                        & 7        & 5        & 4        & 4        & 4        & 3        & 3        & 2        & 2        & 2        & 1        & \\
  \end{tabular}
  \caption{Features selected by CFS for each dataset are denoted with the $\circ$ symbol. \emph{nIssuesByCreator} appeared in only 1 dataset suggesting it is not a good predictor of issue close time, while \emph{nCommitsInProject}, \emph{nIssuesCreatedInProject}, and \emph{nIssuesCreatedInProjectClosed} appeared in 8, suggesting they are clearly important in predicting issue close time.}
  \label{tab:cfs}
\end{table}

Further to the last point, our CFS tool selects
 very few features per project.
 The results of the CFS selection are shown in Table \ref{tab:cfs} (our target class, {\em timeOpen}, is omitted from that table).
In that figure,
columns are sorted according to how many features
were selected within a data set and
rows are sorted according to how many data sets used
an feature.
That figure shows that:
\bi
\item
{\em node} uses all of the
features defined in   Table~\ref{tab:features}.
\item
Most data sets use a small minority of the features:
the median number is 3 and often it is much less;
e.g. {\em cocoon} only uses one feature.
\item
 The
number of selected features is not
associated with the success of the learning. For example,
consider the datasets {\em hive} and {\em cocoon} which make
use of many and few features (respectively), but both achieve poor cross-val results despite using a different number of features.
\ei
From the above, we cannot say that any particular set
of features is always ``best'' (since CFS selects
different features for different data sets).


\section{Threats to Validity}

As with any empirical study, biases can affect the final
results. Therefore, any conclusions made from this work must
be considered with the following issues in mind.

{\em Sampling bias} threatens any classification experiment;
i.e., what matters there may not be true here. For example,
the data sets used were selected by our industrial
partners (since they use and/or enhance these particular
projects).  Even though these data set   covers a large scope
of applications (see Figure \ref{fig:alldata}), they are all open
source systems and many of them are concerned with Big Data applications.

{\em Learner bias:} For building the defect predictors in this
study, we elected to use a single decision tree learner. We chose the this learner since past experience showed it generates
simple models and we were worried about how to
explain our learned models to our industrial partners.
That said, data mining  is a large and active field
and any single study can only use a small subset of the known
classification algorithms.

{\em Goal bias:} We used the same ``N-binaries'' approach
as used in the prior state-of-the-art in this work (the KDP
paper) but whereas   they built $N=24$ different models, we only built
$N=5$ models. Given our current results, that decision seems justified
but further work is required to check how many $N$ time thresholds are most useful.


\section{Conclusions}

Our results support several of the conclusions made by recent results from KDP.
Firstly, we endorse the use of ``N-binary'' learners.
Our explanation for some of the poor results seen in prior
studies is that they were trying to make one model do too much (i.e. predict for too many classes). At least
for predicting issue lifetime, it would seem that
$N$ learners each predicting ``before/after'' for a particular
time threshold performs very well indeed.

Secondly, we endorse KDP's conclusion that it is best to use contextual
features for predicting issue close time. All
the results of this paper, including our excellent cross-project
results, were achieved without reference to static code measures (except \emph{issueCleanedBodyLen}, used in models for 3 projects).
Further,  just a handful of contextual features (see right-hand-side of Table \ref{tab:cfs}), are enough
to predict issue lifetimes.
Like KDP, these results say that issue lifetimes can be characterized
by the temporal pattern of issues and commits associated with the team
working those issues. Our models seem to act like a distance function
that find the closest temporal pattern to that seen in some
current issue. Once that closest pattern is found, we need only
report the average close time for that group of issues and commit
patterns.

That said, we achieve our positive results with three of our own methods. As shown in Table~\ref{tab:comparison}, using the following methods, we achieved
lower false alarm rates and higher precisions than KDP:
\bi
\item
Prior to learning, use CFS to find the best features.
\item
Use simpler learners
than Random Forests. As shown in Figure~\ref{fig:trees}, we can produce very
simple decision tree using C4.5, which are much
simpler to show and explain to business users
than Random Forests. 
\item
The best way to learn issue lifetime predictors is to use data from other projects.
\ei
As evidence for the last point, 
the results of  Table \ref{tab:results} are very clear:
round-robin learning (where the training data comes from
other projects) does better than local learning.
Recall that in that table, we had many data sets where the local learning
failed spectacularly ({\em cocoon, hadoop, hive, kafka, node, ofbiz, qpid}). Yet in each of those cases, the cross-project
results offered effective predictions for issue lifetime.

To end this paper, we repeat a note offered
in the introduction.
The  success  of  cross-project  learning  for  lifetime  prediction was  quite  unexpected. These  results
were achieved without the data adjustments recommended in the defect prediction transfer learning literature~\cite{turhan2009relative,kocaguneli2015transfer,nam2015heterogeneous}.
That is, while transfer learning for some software analytics tasks can be complex, we have discovered that there exist some tasks such as predicting issue lifetimes where it is relatively simple.
Perhaps, in 2017, it is timely to revisit old conclusions
about what tasks work best for what domains.

\section*{Acknowledgments}
This work is funded by the National Science Foundation under REU Grant No. 1559593
and via a faculty gift from the DevOps Analytics team in the 
Cloud Division of IBM, Research Triangle Park, North Carolina.

\bibliographystyle{elsarticle-num}
\balance
\bibliography{refs.bib}
\balance

\end{document}